\documentstyle[aps,prl,epsfig,twocolumn]{revtex}

\begin{document} 
 
\title{Particle dynamics in sheared granular matter} 
 
\author{W. Losert$^{1}$, L. Bocquet$^{2,3}$, T.C. Lubensky$^{2}$, 
and J.P. Gollub$^{1,2}$} 

\address{$^{1}$ Physics Department, Haverford College, Haverford, PA 19041}
\address{$^{2}$ Physics Department, University of Pennsylvania,
Philadelphia, PA 19104}
\address{$^{3}$ Laboratoire de Physique de l'E.N.S. de Lyon, U.M.R. 
C.N.R.S. 5672, 46 All\'ee 
d'Italie, 69364 Lyon Cedex, France}

\date{\today}

\maketitle

\begin{abstract}
The particle dynamics and shear forces of granular
matter in a Couette geometry are determined experimentally.  
The normalized tangential velocity $V(y)$ declines strongly with distance
$y$ from the moving wall,
independent of the shear rate and of the shear dynamics.
Local RMS velocity fluctuations 
$\delta V(y)$
scale with the local velocity gradient to the power $0.4 \pm 0.05$. 
These results agree with a locally Newtonian, 
continuum model, where the granular medium is assumed to behave as a 
liquid with a local temperature $\delta V(y)^2$ and density dependent
viscosity.
\end{abstract}

An important property of granular matter is partial fluidization in response 
to shear stresses~\cite{Nagel96}.  Stationary granular matter can sustain 
normal loads and shear stresses, but if a threshold shear stress is exceeded, 
part of the material starts to flow with properties that appear to
differ from those of a Newtonian fluid. 
Unlike ordinary fluids, granular materials do not exhibit 
intrinsic thermal motion.  Instead, the granular 'temperature', generally
defined as the square of RMS velocity fluctuations $\delta V^2$,
is created by the flow itself.  
As a result, the mean flow and RMS fluctuations are related. 
This fundamental connection has been investigated 
experimentally~\cite{Durian}, but remains poorly understood. 

The flow of sheared granular materials has been investigated 
in the steady state in several experiments by shearing material in a 
Couette cell between
a stationary outer cylinder and a rotating inner 
cylinder~\cite{Schoellmann99,Howell98,Mueth2000}. 
All experiments indicate that the mean particle velocity parallel 
to the shear direction $V(y)$ decreases faster than linearly away from the
inner cylinder.  

The velocity profile in three dimensions was determined by 
Mueth {\it et al.}~\cite{Mueth2000}.  Measurements were carried out both
in the interior of the material using X-ray and NMR techniques, and on 
the bottom surface of the Couette cell by optical imaging.  These 
measurements showed 
that the velocity profile on the bottom surface and in
the interior are the same.  $V(y)$ varies
from nearly Gaussian for kidney shaped particles
to roughly exponential for spherical particles.
Layering of particles is suggested as the cause of this variation.

In previous studies in a planar geometry~\cite{Losert2000}, we have found 
that most of the flow is confined to $5-6$ layers of particles 
close to the shear plane.  
The mean particle velocities during brief slips
of the shearing plate decrease roughly exponentially with distance
away from the moving plate, consistent with the findings in the Couette
geometry.  

The aim of the present experiment is to understand the relationship between
mean velocities,
RMS fluctuations and shear forces of a steady state shear flow.
We have developed a locally Newtonian, continuum model that describes 
the granular medium 
as a liquid with nonuniform temperature and density dependent viscosity.  
The interplay between mean flow and RMS velocity fluctuations
can be understood quantitatively in this context, as we demonstrate. 

In order to accomplish fluidization independent of shear, 
we apply an upward airflow at a variable rate 
through granular matter sheared in a Couette geometry. 
We measure both the mean particle velocities $V(y)$ 
and the velocity fluctuations $\delta V(y)$ on the upper surface
of the granular material.  These should approximate 
particle motion in the interior based on the previous 3D  
measurements described above~\cite{Mueth2000}.
Measurements of shear forces in air fluidized granular matter, 
including a discussion of previous related studies of 
shear forces~\cite{Tardos98},
will be presented elsewhere~\cite{Losert2000b}.

\begin{figure}
 \begin{center}
\epsfig{file=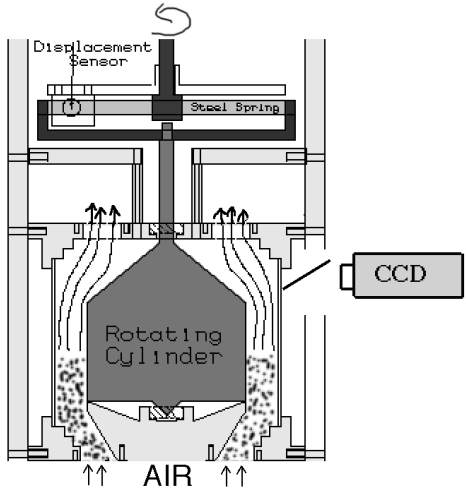,width=\linewidth}
\end{center}
\caption{Experimental setup:  The granular material 
(between two concentric cylinders)
is fluidized by an upward airflow and sheared by rotation of the inner 
cylinder, which is connected to the 
motor through a flexible spring (S). 
Shear forces are determined from the spring 
displacement.  Particle motions in the top layer are measured
through the glass outer cylinder with a fast CCD camera.
 }
\label{exp_setup}
\end{figure}

In the experimental apparatus the granular material ($0.75 {\rm~ mm}$ 
diameter black glass beads) is confined to a $15 {\rm~ mm}$ gap between 
a stationary outer cylinder and a rotating inner cylinder ($r=51 {\rm~ mm}$), 
as shown in Fig~\ref{exp_setup}. 
The inner cylinder is hollow to reduce its inertia and is
coated with a monolayer
of randomly packed glass beads.  The outer cylinder is made of glass and is
coated with a  monolayer of randomly packed glass beads up to the 
height of the top surface,
which allows observation of the top layer of grains through a mirror as
shown in Fig~\ref{exp_setup}.  
To shear the material, the inner cylinder is rotated at a 
variable rate of $0.001 - 1$ Hz. 
The lower $38 {\rm~ mm}$ of the inner cylinder is stationary 
in order to minimize boundary layer effects.
 
The inner cylinder is connected to a microstepping motor via a flexible 
tempered steel spring.  The spring bending is proportional to the applied
 shear force.  We measure the spring 
displacement with a capacitive displacement sensor that is rigidly 
connected to the motor shaft.  This spring configuration allows us 
to measure instantaneous shear forces with excellent dynamic 
range and precision,
and permits both stick-slip dynamics and continuous motion of the 
inner cylinder, depending on parameters.  The trajectories of roughly $100$ 
individual particles in the surface layer are determined with a fast 
CCD camera at $30-1000$ frames/sec.  Particle motion is extracted 
from sequences of $2000$ images with a spatial resolution of $< 0.1$ pixels.  
From the particle tracks we determine average particle velocities $V(y)$ and 
RMS velocity fluctuations $\delta V(y)$ 
as a function of distance $y$ from the rotating inner cylinder. 

The shear dynamics without airflow are found to be very similar to 
the dynamics of a plate sliding across a granular 
layer~\cite{Geminard99}.  
At low shear rates the motion of the inner cylinder 
is intermittent with short, rapid slips, 
and long periods of sticking.  At sufficiently high shear rates 
(or with a stiff spring), steady motion of the inner cylinder is observed.  

Airflow has a strong effect on the shear forces and shear dynamics.  
Without airflow, stick-slip 
motion is observed with a mean shear force of $ 95 {\rm~ N/m^2}$.  
At small airflow, the shear force is reduced, but stick-slip motion persists.
At high airflow rates (with a mean air speed of  $\sim 1.6 {\rm ~ m/s}$ 
through the granular material), shear forces are reduced by a factor of 
4 to $\sim 20 {\rm~ N/m^2}$ and steady sliding motion is observed.
The experiments described below were carried out at large airflow 
just below the 
threshold of the 'bubbling' instability.  For $0.75$ mm diameter glass 
particles no motion on the upper surface is observed at this airflow in the
absence of shear. 
The mean shear stress does not change with shear rate over more 
than two orders of magnitude in the shear rate.
The shear forces  
will be discussed in detail elsewhere~\cite{Losert2000b}.
Here we note that airflow strongly reduces the shear strength of the material
and changes the dynamics from stick-slip motion to steady sliding.

\begin{figure}[h] 
\begin{center}
\epsfig{file=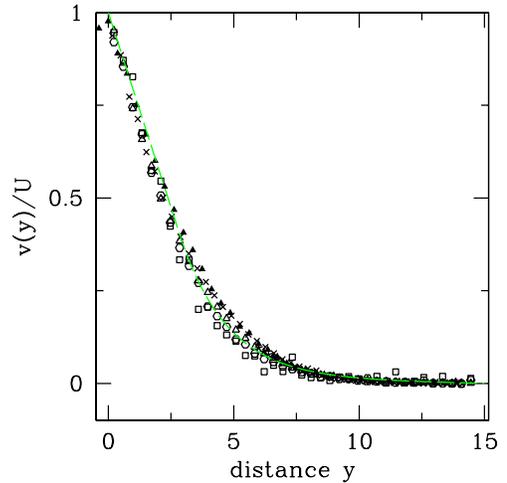,width=6.5cm,height=6.5cm}
\end{center}
\caption{Mean particle velocity (normalized by the shear rate)
 as a function of distance from the inner 
cylinder (in particle diameters). 
The respective shear rates $U$ (in Hz) are : 0.004 (hexagons), 0.04
(squares), 0.01 (open triangles), 0.4 (crosses). The
solid triangle is the velocity profile at $U=0.01$ Hz {\it without} airflow.
The normalized velocity profile is independent of
shear rate and shear dynamics (intermittent or steady motion).
The dashed line is the
solution of eqs. (2) and (3), with $\delta=4.7 d$, $y_w=2.8 d$, and
$\alpha=.4$ ( see text for details).
}
\label{vel_profile}
\end{figure}

The average velocity of particles at the surface of the shear cell is 
shown in Fig.~\ref{vel_profile} as a function of distance $y$ from the 
rotating inner cylinder, for inner wall velocities $U$ ranging from 
$0.004$ to $0.4$ Hz. As found previously, the velocity profile
decays strongly  to zero over a few particle diameters $d$.
The normalized velocity profile is independent of the imposed 
shear rate over at least two orders of magnitude in shear rate.
We also find that without airflow (at  
$U=0.01$ Hz, solid triangles), 
the inner cylinder moves with stick-slip dynamics rather than steady sliding,
but the velocity profile remains unchanged.
   
\begin{figure}[t] \begin{center}
\epsfig{file=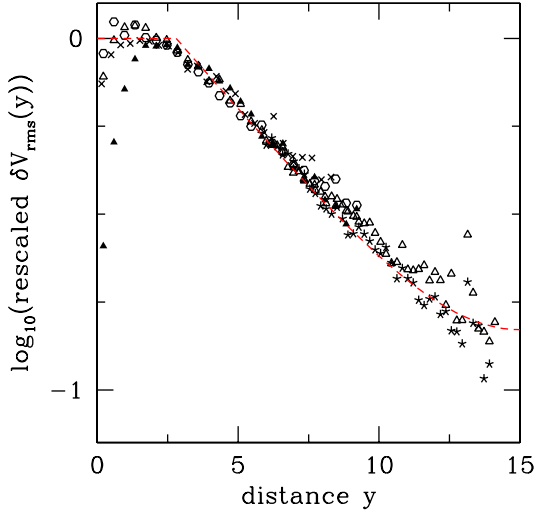,width=7cm,height=7cm}
\end{center}
\caption{RMS velocity fluctuations perpendicular to the shear direction.
Fluctuations decrease roughly exponentially far from the inner cylinder,
but more slowly than the mean flow. The rms fluctuations are rescaled 
(shifted vertically) such
that all epxerimental points fit on the same master curve.
The dashed line is the theoretical
result (see text), with a decay length $\delta=4.7 d$ and a boundary position
$y_w=2.8 d$.}
\label{vel_fluct}
\end{figure}

Figure~\ref{vel_fluct} shows the RMS velocity fluctuations, which
have not been previously measured in a 3D system.
The fluctuations are slightly 
larger in the direction parallel to the mean flow than perpendicular 
to it.  Since parallel fluctuations would 
also include the effect of fluctuations 
in the mean flow, we  show only the perpendicular fluctuations.  
The velocity fluctuations decrease roughly exponentially far from the inner 
cylinder. However, the fluctuations decrease more slowly with $y$ than does 
the average velocity.  

The RMS fluctuation is the key quantity in a flowing granular
material. As already proposed by Reynolds 
\cite{Reynolds}, the system has to dilate in order to allow 
flow, which implies particle motion transverse to the flow direction.
This transverse motion vanishes on average and,
therefore, its RMS fluctuations contain the physically 
relevant information characterizing the flow properties.

In a phenomenological description, it is tempting
to consider these RMS velocity fluctuations as an effective
{\it granular temperature} and consider  
a {\it local} hydrodynamic model as a starting point, as has already been
proposed in the literature
(see e.g. \cite{KT}). Within such a picture the
granular temperature obeys a transport equation characterizing
the balance between ``heat flow'' and dissipation due to
the inelastic collisions :
\begin{equation}
{\partial_{y}} \lambda {\partial_{y}} T
- \epsilon T = 0.
\label{heateq}
\end{equation}
(Note that a local heating term has been omitted, which can be shown
to be valid far from the moving boundary.)
This {\it local} equation introduces
two transport coefficients, the thermal conductivity $\lambda$ and
the energy loss rate $\epsilon$. 
Both $\lambda$ and $\epsilon$ are proportional to the collision
frequency. In the high density limit we consider, $\rho\sim\rho_c$
with $\rho_c$ the density at random close packing, $\lambda$ and $\epsilon$
are proportional
to $P/T^{1/2}$, where $P$ is the isotropic pressure \cite{KT}.
Stress conservation in the $y$ direction implies that
$P$ is constant over the cell. 
From eq. (\ref{heateq}), one can obtain the temperature profile as 
\begin{equation}
T^{1/2}(y)= T_0^{1/2} {\cosh({(H-y)/\delta})\over 
\cosh({H/\delta})}
\end{equation}
where $\delta$, defined as $\delta^2= 2\lambda/\epsilon$, is proportional 
to the bead diameter $d$, and is independent of position and temperature. 
A vanishing heat flux has been assumed at the outer stationary boundary
in agreement with additional experimental measurements for wider shear regions, 
while the temperature at the moving boundary, $T_0$, is introduced 
as a boundary condition at the wall. $T_0$ is created by the flow 
in the first layers close to the wall, so that eq. (\ref{heateq})
is assumed to  apply only for distance $y$ larger than a cutoff
distance $y_w$ from the wall : $y_w$ quantifies the width of the boundary layer
in which temperature is created by the motion of the inner cylinder.
The fit to experimental data yields $y_w=2.8 d$.
The effective cell width $H$ is thus
$H_0-y_w$, where $H_0$ is the wall to wall distance.
As shown in Fig. \ref{vel_fluct}, this hydrodynamic model shows 
good agreement with the experimental data.

\begin{figure}[t] \begin{center}
\epsfig{file=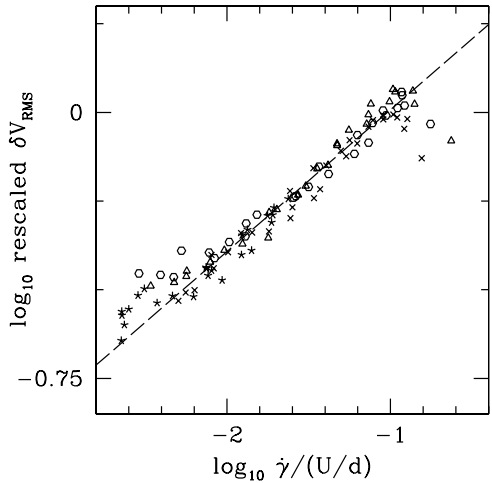,width=6.5cm,height=6.5cm}
\end{center}
 \caption{Connection between the local RMS velocity fluctuations and local
shear rate (same symbols as for Fig. \ref{vel_profile}). 
Local fluctuations are found to increase approximatively as a power law 
of the local velocity gradient, with a power of $0.4$ (dashed line). }
\label{fluct_grad}
\end{figure}

Within the framework of this phenomenological model, the velocity profile 
can be constructed from the constant shear-rate condition, 
$\sigma_{xy}=\eta \dot\gamma=const$,
with $\dot\gamma=dV_x(y)/dy$  the local shear rate and $\eta$  the 
viscosity.
The viscosity is expected to scale with the collision frequency, 
$\eta = \eta_0 P/(\rho_c d^2 T^{1/2})$ (where $d$ is the bead diameter,
$\eta_0$ a dimensionless number and $\rho\sim\rho_c$ has been assumed), so that the
local shear rate obeys $T^{1/2}=\eta_0 P/(\rho_c d^2\sigma_{xy})\dot\gamma$.
This local relationship can be compared to the experimental data 
by plotting the local RMS velocity as a function of the local velocity 
gradient $\dot\gamma$.  As shown in Fig.~\ref{fluct_grad}, a 
remarkable scaling relationship is found between these two local quantities,
\begin{equation}
\delta V_{RMS} \equiv T^{1/2} \sim \dot\gamma ^\alpha
\label{scaling}
\end{equation}
with an exponent $\alpha\simeq 0.4 $. This scaling behaviour is not
in agreement with the simple scaling of the viscosity with the collision
frequency only, for which an exponent of $1$ would have been
measured. 

The actual scaling can be understood if we assume
that the dimensionless coefficient $\eta_0$
contains an additional contribution that 
diverges algebraically,  when the density reaches random-close packing, 
$\rho_c$, i.e. $\eta_0=\tilde\eta_0 (1-\rho/\rho_c)^{-\beta}$. 
The density can be related to the granular temperature through 
the constant isotropic pressure $P$, which can be formally written 
as $P=\rho f(\rho) T$.
The dimensionless function $f(\rho)$ is expected to diverge like
$(1-\rho/\rho_c)^{-1}$ at random close packing \cite{KT}, so that at
constant pressure $P$, in a region
where $\rho \sim \rho_c$, one has $\rho_c-\rho \sim T$. Note that this 
relation is consistent with Reynold's remark that the 
density has to decrease in
order to allow flow, which occurs for non-zero granular temperature.
Together with the constant shear rate condition, these relations imply
a nonlinear algebraic scaling of $\delta V(y)$ with the shear rate,
with an exponent $\alpha=(2\beta+1)^{-1}$. With the experimentally determined
exponent $\alpha=0.4$, we obtain the overall divergence of the viscosity 
with density as $\eta \sim (\rho_c-\rho)^{1.75}$.

Using
the previously obtained temperature profile, Eq. (\ref{scaling})  can
be integrated (with no-slip boundary conditions
at both walls) to give the velocity profile. The
corresponding result is plotted in Fig. \ref{vel_profile} as a dashed
line, indicating good agreement with the experimental profile. 
The calculation also yields a relation between the
boundary temperature $T_0$ and the shear rate $U$, 
as $T_0 \sim U^{\alpha}$ in agreement with the experimental data (not shown).

Finally, we measure granular flow
when the gap between the two cylinders is only 5 particle diameters wide.
In this case,
the inner cylinder jams and can only be moved at high airflow rates
and high rotation rates.  For the narrow gap we find that $V(y)$ decreases
linearly with $y$ and that $\delta V(y)$ is
uniform across the gap at shear rates of $0.1$ rev./sec and $0.02$ rev./sec.
This result is consistent with our model: 
A linear velocity profile is expected for a constant granular temperature.
A constant granular temperature in turn is expected for cell widths smaller 
than the previously defined decay length $\delta$.

The success of the present {\it local} {\it hydrodynamic} model in
describing the experimental data
(even for small driving velocities) calls for a deeper
understanding. The basic ingredient of the present model is the algebraicly
diverging viscosity close to random close packing. 
In a granular material, the ability to flow relies mainly on a
decrease in the local density.  Such a decrease can be caused by the onset of 
RMS velocity fluctuations.  
Lowering  the density allows some particles to move, creating
``fluidized regions'', while other regions remain in clusters at rest. 
The flow properties can be thus formulated in terms of  the local 
fluidization probability, or cluster lifetime.  In a granular
material, these quantities might be estimated by means of a free volume 
estimate, along the lines proposed by Edwards \cite{Edwards} for granular
matter and earlier by Cohen and Turnbull \cite{Cohen} in the context of 
transport in glassy systems. 
These models predict a very strong divergence of the viscosity
close to random close packing, in qualitative agreement with the present
results. More precisely, our data are consistent with an algrebraic  
divergence, which is also expected in supercooled liquids when the 
density moves off the critical density \cite{gotze}.

The model quantitatively describes all aspects of the
experimental results: (i) The RMS velocity fluctuations increase with 
the local shear rate to a given power. (ii) The velocity decreases
strongly away from the moving wall. (iii) The normalized velocity profile
is independent of shear rate and shear dynamics.

More generally, our experimental results indicate 
that there may be a useful analogy 
between the dynamics of granular materials and the behavior of supercooled
liquids close to the glass transition. Such an analogy was proposed
in a recent paper by Liu and Nagel \cite{LiuNagel}.
Our results support this conjecture and indicate the quantitative 
relationship:  
The flow properties reported here are quantitatively predicted 
from a locally Newtonian, continuum model, provided that the
local temperature is identified with $(\delta V)^2$ and that
the viscosity is assumed to diverge as the
density approaches a critical value. 
The specific properties of individual particles only enter
through the few adjustable parameters of the model. Once determined in a
specific geometry, the flow properties in any geometry can be predicted. 
Further experiments are underway to verify this predictive power of the model.

\section{Acknowledgments}
We thank A. Liu, H. Jaeger and C. Bizon for helpful discussions.
Part of this work was supported by the National Science Foundation under 
Grant DMR-9704301.







\end{document}